# A 4/3-ratio approximation for TAP by Deferred Primal-Dual

Guy Kortsarz
Rutgers University, Camden
guyk@camden.rutgers.edu

**Abstract**

We study the *Tree Augmentation Problem (TAP)*. Given a tree $T = (V, E_T)$, and additional set of *links* $E$ on $V \times V$ the goal is to find $F \subseteq E$ such that $T \cup F$ is 2-edge-connected, and $|F|$ is minimum. We give an a 4/3 approximation for TAP approximation, improving the best known approximation 1.393 by F. Cecchetto, V. Traub and R. Zenklusen J. ACM 2025 [44]. The running time is $O(m \cdot \sqrt{n})$ using the algorithm of [34],[45] for maximum matching. Faster than [44], [42], [41] as our algorithm does not enumerate structures of size $\Theta(1/\epsilon)$. We introduce the *deferred primal-dual* technique where cuts are not disjoint in different stages (a links can cover later cuts) in addition to new insight into [31]. Another new technical idea is finding *in advance* links whose inclusion imply the optimum *is larger*. We assign "penalty" to such links. This allows to simplify and discard several of the difficult ideas of [31]. We change the definition of the subtrees we cover to allow primal-dual/local-ratio steps [25],[3],[2] [4] in the algorithm.

# 1 Introduction

We study the following problem:

| Tree Augmentation Problem (TAP) |
|---|
| *Input:* A tree $T = (V, E_T)$ and an additional set of *link* $E$ on $V \times V$. |
| *Output:* A a link set $F \subseteq E$ such that $T + F$ is 2-edge-connected and $\|F\|$ is minimum. |

Our main result is:

**Theorem 1.** *TAP admits a* $4/3$*-ratio algorithm whose running time is* $O(m \cdot \sqrt{n})$*, with* $m$ *the number of links in the tree and* $n$ *the number of nodes.*

This improves the 1.393 approximation by F. Cecchetto, V. Traub and R. Zenklusen, J. ACM 2025 [44]. See also [42] (SODA 2024) and [41] (FOCS 2021). The algorithms is faster than [44], [42], [41], as we do not enumerate structures of size $\Theta(1/\epsilon)$ and it is dominated by $O(m \cdot \sqrt{n})$ run time maximum matching computation using [34]. See a more recent simplification by [45]. The running time is discussed in Section A in the appendix.

# 2 Previous work

The weighted version, WTAP is: given a tree $T = (V, E_T)$ and an additional set of *link* $E$ on $V \times V$ with weight $w : E \mapsto Z^+$, find a minimum weight subsets $F$ so that $G + F$ is 2-edge connected and $F$ has minimum weight. A ratio 2 approximation for for WTAP is given in [22]. One way to get ratio 2 is by replacing every link $ab$ by $ax, bx$ so that $x = lca(a, b)$. The weight of the optimum at most doubles. The problem becomes polynomial since after this replacement the LP is totally unimodular (see [16]). WTAP is equivalent to increasing connectivity from $k$ to $k + 1$, *for an odd* $k$ since in this case the Cactus Structure for minimum cuts is a tree. See [18]. It is also equivalent to covering a laminar family (see [13]).

**Previous Work on TAP:** Approximating TAP within 2 is folklore. Over time, the approximation ratio has been progressively improved: 1.9 [35], 1.8 [20], 1.75 [32], 1.5 [31], 1.458 [27], 1.393 [43] [44]. The latter gives hypergraphic LP integrality gap. This is the best ratio known prior to our paper.

The integrality gap for the *basic LP* for WTAP is at most $2 - 2/15$ [38] and at least 1.5 [14]. For constant maximum weight, integrality of $3/2$ is known [21], using Chvátal-Gomory Cuts.

**Previous work on WTAP:** Approximation ratio for WTAP progressively improved: 2 [22], 1.9695 for constant maximum weight, [1], $3/2 + \epsilon$ for constant weights [21], $12/7$ approximation for $O(\log n)$ weights [38], a $1 + \ln 2$ approximation for WTAP for bounded diameter trees [15]. The first paper to break the ratio 2 for WTAP is [7] by F. Grandoni, A. J. Ameli and V. Traub. The approximation ratio is slightly better than 1.91. The method used is a reduction to the Steiner tree problem. This is later improve in [41] to 1.7, using the *relative greedy technique,* and to $1.5+\epsilon$, [42]. The journal version is [44].

The *Weighted Connectivity Augmentation Problem* is to increases the connectivity of a graph connected to the $k$-edge from $k$ to $k+1$ *for even* $k$ (recall the $k$ is odd case, is equivalent to WTAP). The problem is equivalent to augmenting connectivity in a *cactus tree.* This is a graph where every edge belongs to at most one simple cycle. This graph is *series-parallel*, namely, has *treewidth* 2 (see [6]). It is also equivalent to increasing the connectivity of a a rooted cycle [43]. The decomposition theorems, [43], includes *cross edges* between components, which do not exist in WTAP.

The 1.91 approximation of [7] is for WCAP. The approximation for WTAP is a byproduct. In [43], V. Traub and R. Zenklusen give a $1.5+\epsilon$ approximation for the problem.

**LTL-TAP:** In this problem links are only between leaves. The first approximation better than 2 was 17/12 [15]. This was improved in [9] to 1.29.

TAP is approximated by lift and project and Lasserre sum of squares in [11],[12]. For papers on related subjects, see, [39], [28], [36], [23], [10],[37], [26].

**Hardness:** The problem is proved to be NP-hard in [22]. In [13], the NP-Hardness of TAP is extended to LTL-TAP. APX-hardness for LTL-TAP is given in [30]. The inapproximability is around $1+1/900$ [30]. See a discussion of that and future directions in Section B in the appendix.

## 2.1 Our conceptual contribution: The Deferred Primal-Dual algorithm

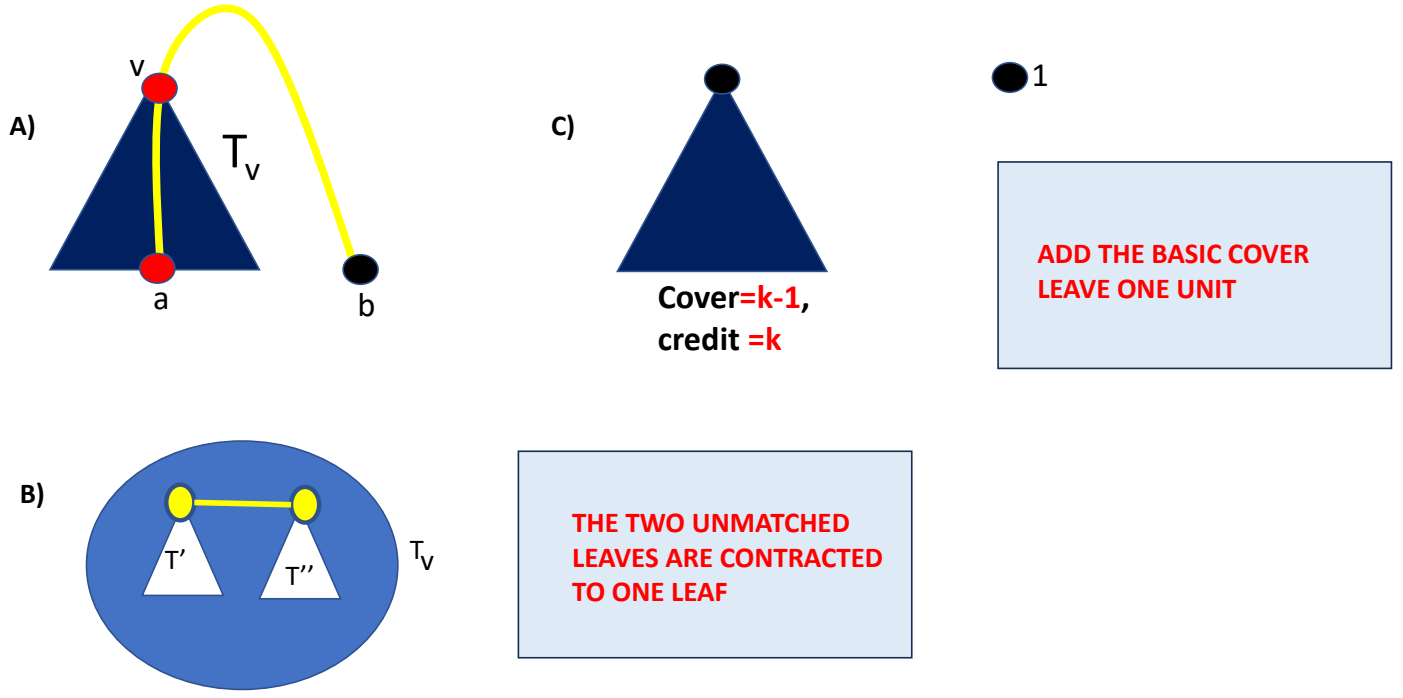

Figure 1: Links that cover leaves outside covered trees

We came to a new understanding of the ideas of [31]. We change the definition of the trees we cover to allow *primal-dual/local-ratio steps in the algorithm* for subtrees with *arbitrary number of leaves*. See [25],[3],[2]. Specifically, we get *disjointness of the optimum* for arbitrarily large trees. The part in the algorithm *after a matching is computed* uses a basic primal-dual/local-ratio step as in [25],[3],[2]. In Section 10 we give a primal-dual interpretation of the of the algorithm after a matching is found. Because we first find a matching, our paper *does not yet give integrality gap* (at least not for the natural LP). We leave this open.

The major conceptual difference from [25] (for example) is that in [25], dual variables (of sets) are increased until spheres meet and are contracted into one big sets. This gives new violates super-sets. *The links (edges) previously added, cannot cover the new cuts.* Our algorithm covers trees called *semi-closed* defined below. Links covering (for example) leaves in the tree *also cover edges outside the tree.* See Figure 1 A). In the figure we see one of the subtrees $T_v$ covered by a the algorithm. A link covering a leaf $a$ in $T_v$, also covers a leaf $b$ outside $T_v$. This happens until we find a *leaves-stems-closed* tree (defined later). *Only then disjointness in the optimum applies.* Hence, the primal-dual step is *deferred* (delayed) until the leaf-closed tree moment.

The way we overcome non-disjointness is using invariants as in [31] (see below). The moral of the story is that *it is possible to add non-disjointness in the primal-dual algorithm, if the disjointness is limited.* We believe that deferred primal-dual algorithm (with the benefit of how we understand it now) will have further applications in the future for other connectivity problems.

**Other conceptual ideas:** We list some of the new techniques.

1. When computing a certain optimal matching, we give penalty on links, by proving *in advance*, these links imply the optimum is larger. This allows us to simplify and remove complex notions from [31] such as *dangerous links.*
2. Our matching is one that *saturates*, namely, has degree 1 leaves and also special nodes called *stems.* This is unlike [31] (there it was an edge cover of leaves).
3. We use special gadget in a non-trivial way in order to get a special type of solution called *shadows-minimal.*
4. The cases analysis is new and contains complex cases.

# 3 Preliminaries

## 3.1 Definitions

We designate a node $r$ to be a root. This defines the standard ancestors/descendants relation.

**Definition 2.** *The subtree $T_v$, is the tree with $v$ and all its descendants. For a link $xy$, let $P_{xy}$ be the path between $x, y$ in the tree. Let $E_{AB}$ be the links with one endpoint in $A$ and the other in $B$.*

In Part A) of the Figure 2, $T_v$ contains $v, x, w, u, c, d$. In part A), $P_{aw}$ has $a, s, r, v, x, w$. $E_{LL} = \{ab, cd, cw\}$

**Definition 3.** Contracting *a link $xy$ means removing the nodes and edges of $P_{xy}$ except for the least common ancestor (lca) of the nodes on the path. A new node created by contracting links is called a* compound node.

Part B) of the figure describes the contraction of $wc$. $w$, $c$ and $u$ enter $x$ due to such an operation. $x$ becomes a *compound node.*

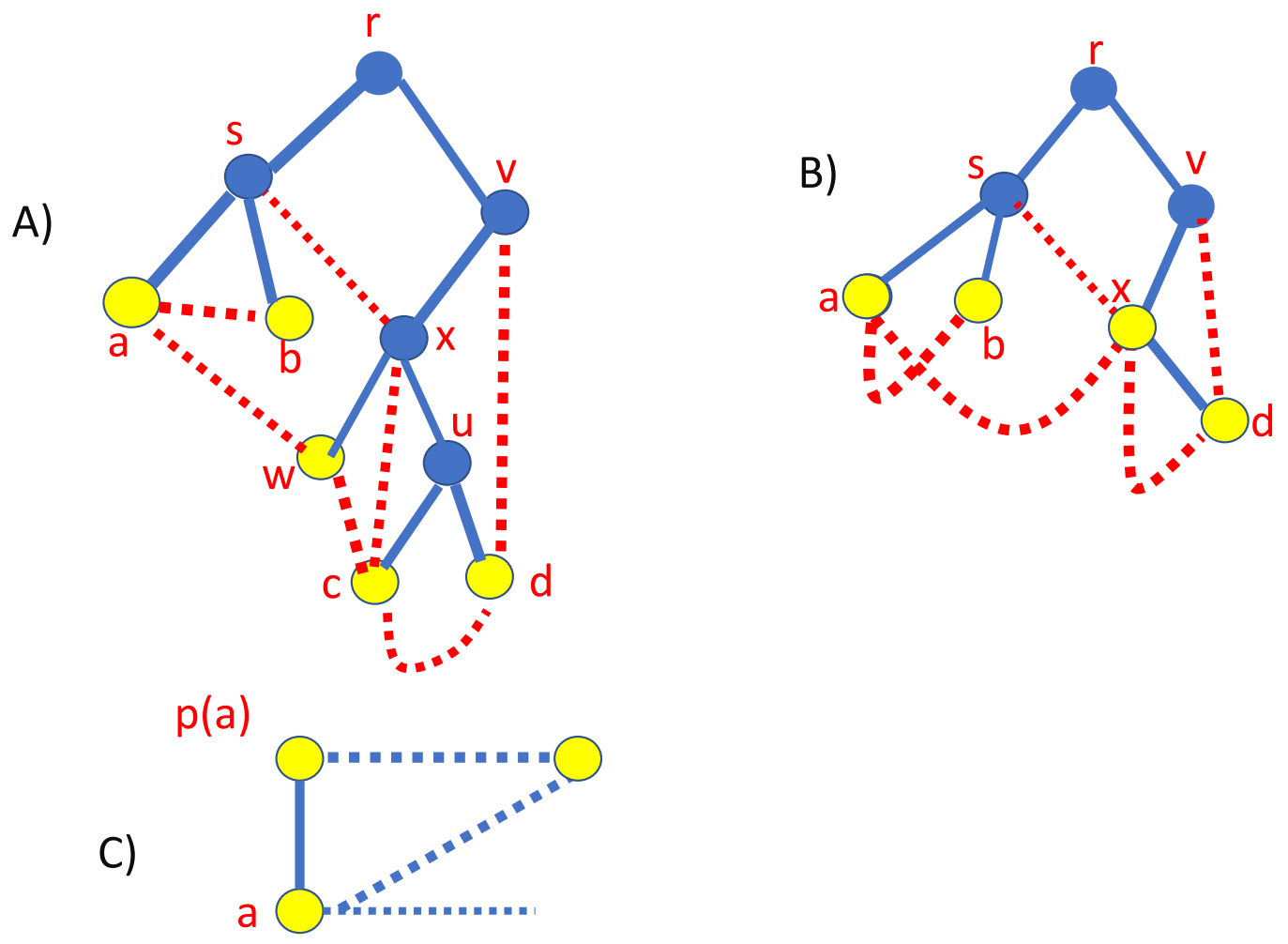

Figure 2: Key concepts

**Definition 4.** [35] **Shadow Completion** *A link cd is a shadow of a link ab if $P_{cd}$ is strict subpath of $P_{ab}$.* We assume that all shadows had been added to the instance.

Shadows in terms of the Set-Cover problem, (see [24]), are a *strict subsets* of an original set. If a shadow is used it can be replaced by the original link. Hence, the addition of shadows does not reduce the optimum.

**Definition 5.** [35] *$F$ is shadows-minimal optimal solution, if $F$ is of minimum size and replacing a link of $F$ by a shadow, renders $F$ infeasible.* [1]

In Figure 2, A) an example of a shadows-minimal solution is is $F = \{aw, bs, cu, dv\}$. $bs$ exists as a shadow of $ab$. $cu$ exists as a shadow of $cd$.

*Covering a node:* For $v \neq r$, covering $v$ is short for covering the parent edge $vp(v)$ of $v$, in $T$. In Figure 2 Part A) $xs$ covers $v$. $dv$ covers $d, u, x$. In order to cover a non-root node $u$, there needs to be a link $xy$ so that, without loss of generality, $x \in T_v$ and $y \notin T_v$. Note that once $T_v$ is contracted, the links covering $vp(v)$ are "shortcut" and now emanate from $v$. The above poses TAP as a Set Cover instance (see [24]).

**Claim 6.** [35] *If $F$ is shadows-minimal, $deg_F(a) = 1, \ \forall a \in L$.*

The claim is illustrated in Figure 2 Part B). A second link to the leaf is replaced by the shadow to the parent.

**Convention:** At any point, all structures depend on $Q$, the links added so far to the partial solution. We omit $Q$ from the notation.

**Definition 7.** [35] **Trees closed and open with respect to nodes.** *We say that $a \in T_v$ is $T_v$-closed, or $T_v$ is $a$-closed if for every link $ax$, $x \in T_v$ as well. Otherwise $a$ is $T_v$ open or $T_v$ is $a$-open.*

[1] In [44], [8], the authors used the term *"every edge is covered once"*.

In Figure 2, $T_v$ is $\{c,d\}$-closed, but is not $w$-closed or $x$-closed because of the links $xs$, $aw$.

**Definition 8.** *Let $L_v = L \cap T_v$ (the leaves inside $T_v$). A tree is $T_v$* leaf-closed *if it is closed for all leaves in $L_v$. A tree is minimally-leaf-closed if it is leaf-closed and no proper subtree of $T_v$ is not leaf-closed.*

Note that $T_v$ of Figure 2 is not leaf-closed because of the link $wa$. Only the full tree $T = T_r$ is leaf-closed. Hence, $T = T_r$ is minimally leaf-closed.

The following is well defined *only* after shadow completion.

**Definition 9.** *The up link of a node $v$ is the link $vx$ to an ancestor $x$ so that $x$ is closest to the root $r$ after shadow completion. This link is denoted $up(v)$. Let $up(U) = \{up(v) \mid v \in U\}$.*

Example: In Figure 2, $up(c) = cx$. $up(a) = ar$. $up(d) = dv$. $up(b) = bs$. $up(w) = wr$.

The reason minimally leaf-closed trees are useful, is that they are covered by their up links.

**Claim 10.** [35] **The Cover Claim.** *For a minimally leaf-closed tree $T_v$ with a set $L_v$ of leaves $up(L_v)$ covers $T_v$*

*Proof.* For the sake of contradiction, assume an edge from $x$ to its parent $p(x)$ is not covered. This implies there are no links from a leaf in $T_x$ to $T \setminus T_x$. This implies $T_x$ is leaf-closed. This contradicts the minimality of $T_v$. □

**Definition 11.** *For the rest of the paper, let $F$ denote an arbitrary* shadows-minimal *optimal solution for the TAP instance at hand.*

**Definition 12.** *Fix matching $M$ on the leaves (to be defined later). Fix $U$ as the leaves unmatched by $M$ )$U = L \setminus V(M)$).*

## 3.2 Basic definitions

**Definition 13.** **Semi-closed** *trees with respect to a matching $M$ on the leaves is defined algorithmically as follows:*

1. *Contracts $M$ to get $T/M$*
2. *Find a minimally leaf-closed subtree $T_v/M$ of $T/M$.*
3. *De-contract $M_v = M \cap T_v$ to get $T_v$*

Consider Figure 2. Say, for example for $M = \{wc\}$. Then, $T/M = T/wc$ is computed. See Part B) of the figure. Then, a minimally leaf-closed subtree of $T/M$ is computed. Note that $T_v/M$ is leaf-closed subtree of $T/M$. In addition $T_x$ is not leaf-closed, because of the link $dv$. This implies $T_v/M$ is minimally leaf-closed subtree of $T/M$. Hence, $T_v$ is a semi-closed tree. However, $T_v$ *is not leaf closed*, and in particular, not $w$-closed. The reason is that $w$ was "inside" $x$ when the $T_v/cw$ had been computed. Hence, it was not a leaf. The following is similar to [31].

**Definition 14. Basic cover.** *The basic cover is* $B(T_v) = M_v \cup up(U_v)$.

**Corollary 14.1. The basic cover corollary:** $M_v \cup up(U_v)$ *covers* $T_v$.

*Proof.* The claim holds because $T_v/M_v$ is minimally leaf-closed and due to Claim 10 (the Cover Claim). □

The basic cover of $T_v$ for $M = \{wc\}$ in Figure 2 is $\{wc, dv\}$.

**Definition 15.** *Let* $cover(T_v) = |B(T_v)|$ *be the number of links in the basic cover of the tree* $T_v$.

**Properties of semi-closed trees**.

**Claim 16.**

1. *A semi-closed tree is closed for unmatched leaves*
2. *A semi-closed tree might not be closed for matched leaves*

*Proof.* Let $a$ be an unmatched leaf in the semi-closed tree, The leaf $a$ might be compound or original leaf. As $a$ is not matched, $a$ is a leaf in $T/M$ (in $T$ after $M$ is contracted). By definition, a semi-closed tree $T_v$ is leaf-closed and therefore closed for $a$. □

# 4 The difficulty: Stems and twin links

Let $U$ be the leaves not matched by $M$. We shall later prove for the specific matching we find: $4 \cdot |M|/3 + |U| \leq 4 \cdot |F|/3$ (recall that $F$ is the optimum shadows-minimal solution). We assign links in $M$ 4/3 units of the lower bound and every leaf in $U$, a unit of the lower bound. By the above inequality, this assigns at most $4 \cdot |F|/3$ units of credit. We iteratively cover semi-closed trees $T_v$ making $v$ a *new leaf.* The following invariants help deal with the non-disjointness.

**Leaf invariant:** Every leaf (compound or not) owns a unit of the lower bound.

**The link in $M$ invariant:** Every link in $M$ owes 4/3 units of the lower bound.

**The independence invariant:** Unmatched leaves do not share links.

If two unmatched leaves share a link, it is added to our solution (denoted by $Q$) and one unit of the lower bound pays for this link and the other is left on the new compound node. See Figure 1, Part B).

The problematic structure is the following.

**Definition 17. Stems:** *A* stem $s$ *s defined by a link* $ab$ *whose contraction creates a leaf. In the case of* $s$. *The link* $ab$ *is a twin link.*

The nodes $s$ and $u$ in Figure 2 are stems. The twin links are $ab$ and $cd$.

**Definition 18.** **$M$-Activated stems.** *A stem* $s$ *is* $M$-activated *if* $ab \in M$ *for the twin link* $ab$ *of* $s$.

**Example:** In Figure 2, $s$ (respectively, $u$) is activated if $ab \in M$ (respectively, $cd \in M$).

The difficulty is when the contraction of a twin link creates a leaf at $s$. While old leaves have a unit of the lower bound, *new leaves do not have that.* We need a stronger lower bound to be able to generate enough lower bound credit, to be able to have an extra unit of credit at new leaves.

## 4.1 The initial lower Bound

Let $F$ be a shadows-minimal optimum solution.

Let $X = V \setminus L$.

**Claim 19. The initial Lower Bound Claim.**

$$4 \cdot |M_F|/3 + |U_F| + \sum_{e \in X} deg_F(x)/3 \leq 4 \cdot |F|/3$$

*Proof.* Place 4/3 units of credit at every link in $F$. This gives $4 \cdot |F|/3$ units of credit. For every link $\ell x$ $\ell \in L$, $x \in X$, *send* 1/3 of the 4/3 credit on $\ell x$ to $x \notin L$. This gives the lower bound $4 \cdot |M_F|/3 + |U_F| + \sum_{x \in X} deg_F(x)/3 \leq 4|F|/3$. □

The crucial term is $\sum_{x \in X} deg_F(x)/3$. This leads to the concept of *golden tickets.*

# 5 Golden tickets

The idea is due to [31]. Informally, a *golden ticket* is *a proof a non-leaf is touched by a link in the optimum $F$.* This gives an increase of 1/3 in the lower bound as can be seen by the initial lower bound. Formally:

**Definition 20.** *We say that $x \in X = V \setminus L$ owns $i$* golden tickets, *if $deg_F(x) = i$.*

The difficulty here is that in order to get an unused unit of the lower bound, *three golden tickets are required.* (in [31] two were enough because every golden ticket was worth 1/2).

We later prove for certain links $e \in F$, their inclusion in $F$ implies that $gt_F(M_F) = \sum_{u \in X} deg_F(u) = i > 0$. We only prove $i \in \{1, 2\}$

**Definition 21. Golden ticket function on links.** *We say that $ab$* gives $i$ golden tickets *if $ab \in F$* implies *$\sum_{x \in X} deg_F(x) = i$.*

As we shall see, certain subtrees imply golden tickets. We define such subtree as *maximal for containment.* This implies that if $\{e_i\}$ is a collection of links and $e_i$ implies $j_i$ golden tickets, the inclusion of all of $\{e_i\}$ in $F$ implies $\sum_i j_i$ golden tickets.

**Definition 22. The Golden Ticket Function on links.** *Let $gt$ be the function from $E$ to $\{0, 1, 2\}$ so that $gt(e) = i$ iff $e \in F$ implies $i$ golden-tickets. Additionally, for a set $\{e_i\}$, each $e_i$ implying $j_i$ golden tickets, $\{e_i\} subseteq F$, implies $\sum_i j_i$ golden tickets.*

The most basic golden ticket is associated with a twin link. If it is activated by $M$, $deg_F(s) = 1$ for the stem $s$. This follows from the shadow-minimality of $F$. This gives extra $1/3$ credit in the initial lower bound because it contributes $1/3$ to $\sum_{x \in X} deg_F(x)/3$.

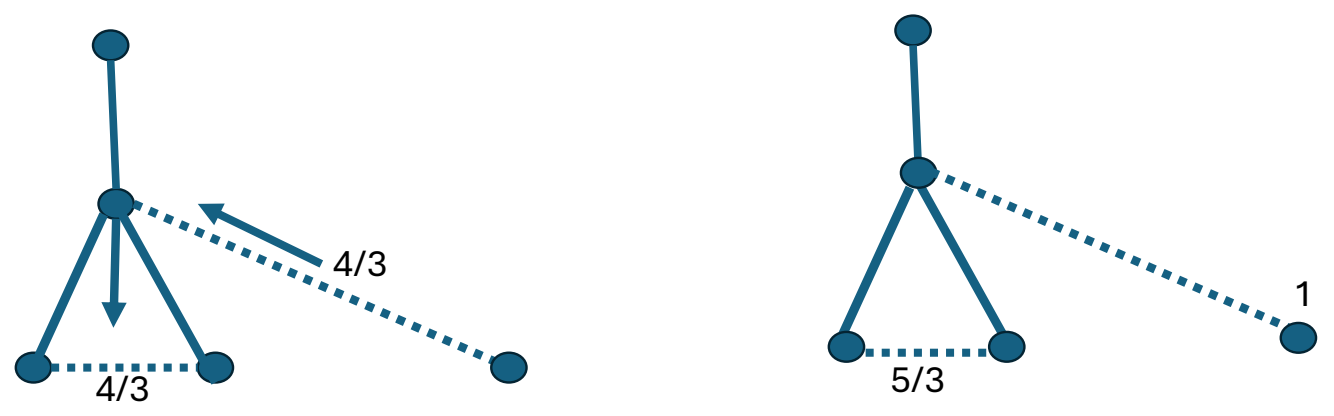

Figure 3: Borrow tickets

**Borrowing Golden Tickets in advance:** Some golden tickets are claimed in advance as penalty on links. A prime example is the case of $M$-activated stems. This is depicted in Figure 3. Normally, we move a $1/3$ of the $cs$ credit to $s$. But in this case, the credit at $s$ is *taken in advance* and is *placed at the twin link* $ab$. This is the *penalty* for choosing a twin link. Usually, links weigh $4/3$ lower bound credit. Namely, its weight is 1 but we multiply by $4/3$ since we aim for $4/3$ approximation. However, borrowing credit in advance for twin links implies the twin weighs $4/3 + 1/3 = 5/3$.

**Fact 1:** Stems do not own golden tickets.

**Claim 23. Tree closed for activated stems claim.** *A semi-closed tree is closed for activates stems*

*Proof.* Let the twin link be $ab$. It was contracted before $T_v/M$ was computed. Hence, it was a leaf. The claim follows because $T_v/M$ is a leaf-closed subtree of $T/M$. □

This means that if we prove that in $F$ there is a link touching a stem $s$, this does not give a golden ticket because it was taken in advance and placed at the twin link. This complicates the analysis. Also, to have disjointness of the optimum, trees have to be *leaf-stem closed*, Since this is the only links used in the lower bound. This creates further complications.

Now, we discuss more "complicated" golden tickets. They require more complex computation. We show in Section A, these golden tickets can be identified in linear time using the algorithm of [40]. Trees as defined below are *maximal for containment.* Hence, node disjoint.

## 5.1 Additional golden tickets

All trees discussed below are assumed to be *maximal for containment* hence node disjoint. This means the tickets are disjoint. We do not further mention this.

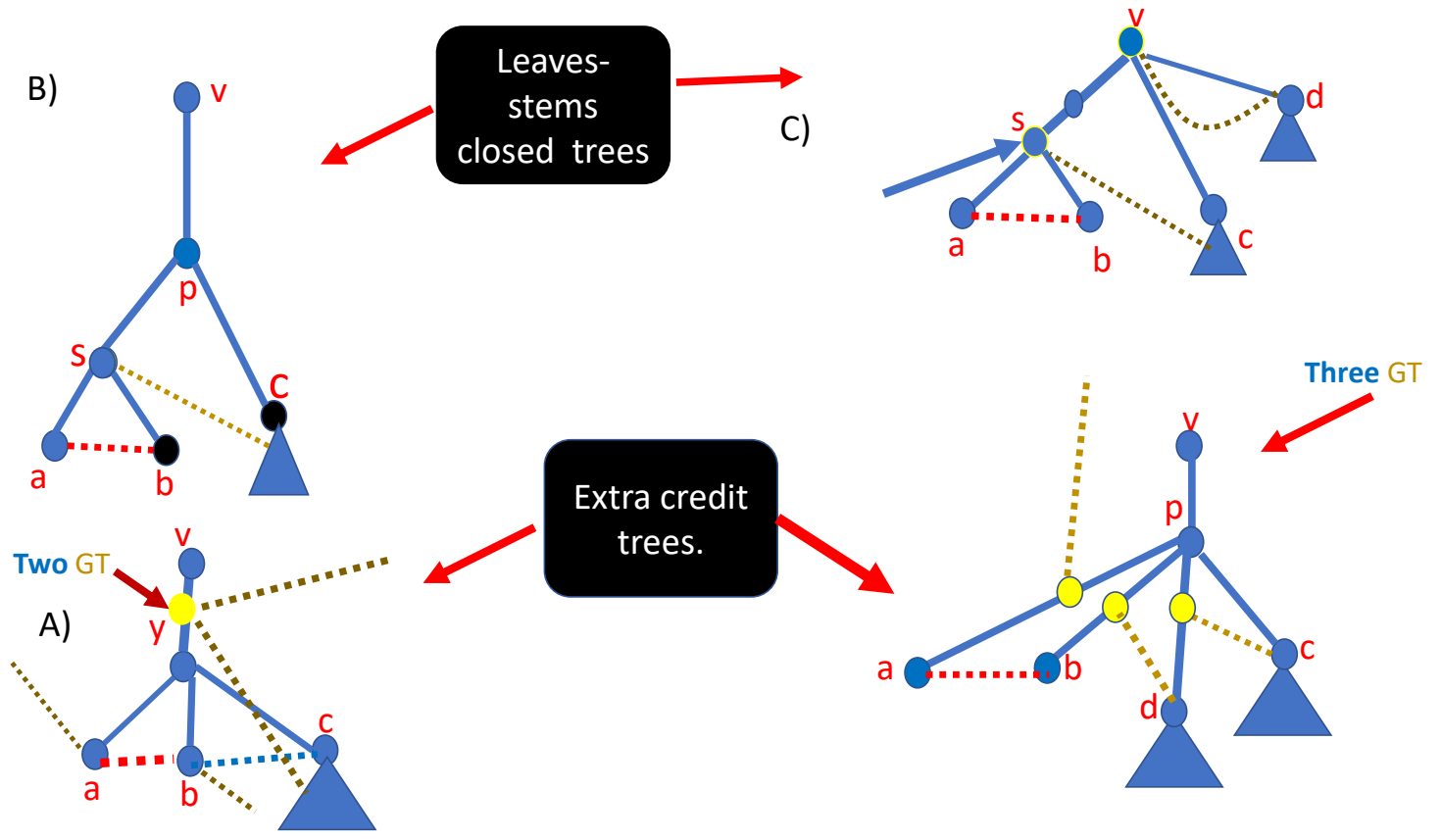

Figure 4: Links that imply golden tickets

In the discussion below we start with Figure 4, the left side of Part A). The assumptions are:

1. $T_v$ has exactly 3 leaves.
2. $T_v$ has no stems
3. $T_v$ is closed for all the nodes in $T_c$.

Note that $c$ might be compound. Under this we proved:

**Claim 24.** *$gt(ab) = 2$, namely, if $ab \in F$, two nodes in $T_v \cap X$ are touched by links of $F$*

*Proof.* Consider the link that covers $c$. By assumption, $ab \in F$. Thus, since $a, b$ have degree 1 in $F$ (they are original leaves) $a, b$ are not linked to $c$ in $F$. Since there are no stems, a node in $V \setminus (S \cup L)$ is touched by a link in $F$, giving a golden ticket.

Now, consider the way $v$ is covered. As $ab \in F$, and because $T_v$ has no stem, no node in $L \cup S$ can cover the $v$, giving a second golden ticket. □

## 5.2 $|L_v| \geq 4$, $|M_v| = 1$ and no stems

Here, we follows Figure 4 Part A) right side. We assume:

1. The tree has no stems
2. $M_v = \{ab\}$
3. The tree has at least two more leaves $c, d$.
4. The tree is closed for all $T_c$ and $T_d$ and similarly, for all other unmatched leaves.
5. There are no links between trees of unmatched leaves (including $T_c$ and $T_d$.

**Claim 25.** $g_F t(ab) \geq 2$

*Proof.* $deg_F(a) = deg_F(b) = 1$ and $ab, cs \in F$. And, the trees has no stems. tickets for covering $c$ and $d$. As the figure shows, an additional ticket can be claimed for covering the root. □

We later see (when we use the claim) this case is not tight. However it might not exist. The case of three leaves is tight.

Other links get $gt(e) = 0$. Note that $F_{LL}$ is a matching $M_F$. This follows from shadow-minimality. We have assigned tickets to links in $M_F$.

**Corollary 25.1. The Golden Ticket Corollary.** $gt(M_F) = \sum_e gt(e) \leq \sum_{u \notin L} deg_F(u)$.

The upper bound might be strict. There might be links in $F_{XX}$. However, it seems that clamming the existence of such links is hard to prove.

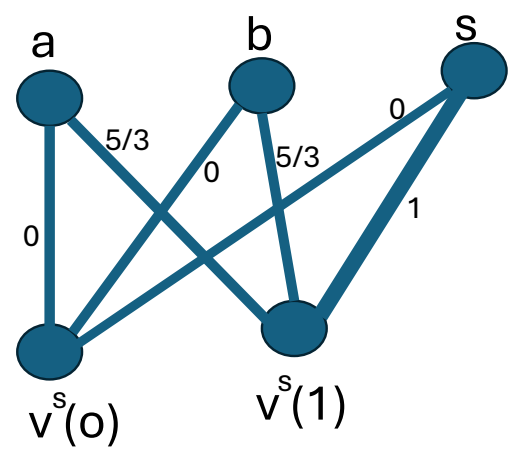

Figure 5: Gadget for twin links

# 6 The Matching computed

We compute a matching on $(S \cup L) \times (S \cup L)$ with $L$ the leaves and $S$ the stems. While we allow $S \times S$ links, in the analysis we assume the worst case; Such links do not exist. Here is the matching we compute:

1. Add a node $x_L$ and links from $L$ to $x_L$ of weight 1. This simulates matching leaves to $X$.

2. Give every other links in $(S \cup L) \times (S \cup L)$, except twin links, weight $w(e) = 4/3 + gt(e)/3$.

3. For every twin links construct the gadget describe in Figure 5. Add two nodes $v^s(o)$ and $v^s(1)$. Let the stem be $s$ and the twins be $a, b$. Add links from $a, b, s$ to $v^s(0)$ of weight 0. Add links of weight 5/3 from $a, b$ to $v^s(1)$. Add a link f weight 1 from $s$ to $v^s(1)$.

Under this, we find a minimum weight matching *saturating* all leaves and stems in the tree. Namely, for leaves and stems the degree must be 1. $v^s(0), v^s(1)$ nodes are allowed to have degree 0 but are not allowed to have degree more than 1. This problem is solved by a reduction to Minimum Weight Perfect Matching. The latter is solved in time $O(m\sqrt{n})$ by the algorithm of [34], [45]. The reduction is given in Section A in the appendix.

Let $\tilde{M}$ be the matching resulting. To get $M$, we remove non leaf-to-leaf links from $\tilde{M}$.

**Remark:** The links of $\tilde{M} \setminus M$ are important. In particular, if a stem $s$ is activated, by the properties of the gadget, $s$ is linked to a node outside $T_s$. And not to $x_L$. To a node in $L \cup S$. We shall assume this node is a leaf $c(s)$ as this is the worst case.

**Claim 26.** *The above gadget simulates choosing a shadows minimal solution so that a twin link weigh* $5/3$*, and having one leaf matched in the gadget weight* $1$ *for the unmatched leaf, and having two leaves matched outside at weight* $0$ *for the stem (as covering the stem in this case should give cost* $0$*)*

*Proof.* The case both $a, b$ are linked to $v^s(0), v^s(1)$ corresponds to selecting the twin link. Note that in this case the 5/3 units of cost are invoked, as should be. In this case $s$ must be linked outside. This simulates a shadows-minimal solution. The case both $a, b$ are linked outside allows $s$ to be linked with weight 0 (as should be when both twins are linked outside). The case one leaf, say $a$ and $s$ are linked in the gadget, invokes a weight of 1 for the unmatched leaf. As should be. □

Hence, we get a corollary

**Corollary 26.1.** *If a stem is* $M$*-activated, then in* $\tilde{M}$*,* $s$ *is linked to some leaf* $c = c(s) \in U$*.*

## 6.1 The main lower bound

**Lemma 27. The Lower Bound Lemma** $w(M) + |U| = 4|M|/3 + |U| + gt(M)/3 \leq 4 \cdot |F|/3$

*Proof.* We observe that, as explained in Figure 3, every node linked to a stem owns a unit of the lower bound and belongs to $U = L - V(M)$. This is true for nodes linked to $x_L$ as well. Thus, $w(\tilde{M}) = 4 \cdot |M|/3 + |U| + gt(M)/3$. Similarly, define the quantities $M_F$ as the matching of $F$ on the leaves, and $U_F$ be the leaves leaf matched to $S \cup X$. We get that $w(\tilde{M}_F) = |M_F| + |U_F| + gt(M_F)/3$.

$w(\tilde{M}) = 4 \cdot |M|/3 + |U| + gt(M)/3 \leq w(\tilde{F}) = 4|M_F|/3 + |U_F| + gt(M_F)/3$ since $\tilde{M}$ has minimum weight. By the Golden Tickets Claim, $gt(M_F)/3 \leq \sum_{x \in X} deg_F(x)/3$. Combining the this, and the initial lower bound, the claim follows. □

## 6.2 Extra credit trees

The credit is distributed with $4/3 + gt(e)/3$ placed at links in $M$ and 1 placed at $U$ not matched by $M$. By the main lower bound, this is at most $4|F|/3$.

**Definition 28.** *Let* $credit(T_v) = 4|M_v|/3 + |U_v| + gt(T_v)$ *be the portion of the lower bound internal to* $T_v$*. Let* $cover(T_v) = |B(T_v)|$ *be the size of the basic cover of* $T_v$*. We say that* $T_v$ *is an* extra credit tree *if* $credit(T_v) = cover(T_v) + 1$*.*

Intuitively, an extra credit tree allows taking the basic cover, covering $T_v$ and leaving a unit of the lower bound on $v$. See Figure 1 Part C). This keeps all invariants. If we can consistently do that, we can cover $T$. We do not spend more than the lower bound, and the 4/3 approximation follows.

For example, if $|M_v| \geq 3$, since taking the links requires one unit of the lower bound but they have 4/3 credit, the tree is an extra credit tree. Hence we may assume:

**Assumption 1:** $|M_v| \leq 2$

**Removing certain stems at the start of the algorithm:** The first step in the algorithm is as follows. Say that $s$ is a stem with twins $a, b$. In addition, at least one of $a, b$, say $a$, is *joined to* $x_L \cup S$. By our credit scheme, $a$ owns a unit of the lower bound.

Then we add link $ab$ to the solution and keep the invariants. We denote the new node by $b$. If $b$ was matched to some leaf $c$, we keep the link $bc$ in the solution with 4/3 credit. Else, ($b$ is matched to a node in $x_L \cup S$) we leave he unit of the lower bound at $b$. This maintaining the invariants.

This gives the following assumption:

**Assumption 2:** For a stem $s$, both twins are matched to a leaf. Either $ab \in M$ or both $a$ and $b$ are matched to two different leaves $c, d$.

**Corollary 28.1.** *If the tree* $T_v$ *has* $|M_v| = 1$ *and a stem, then the stem is activated. Also, the tree is closed for the stem and twins*

*Proof.* If the stem is not activated, by Assumption 2, $|M_v| \geq 2$. Also, the stem and twins were inside a leaf when $T_v/M$ was computed. □

**Lemma 29. Main Lemma:** *Under Assumption 2, there is an algorithm that finds a tree of one of two types:*

1. **Extra credit tree:** $credit(T_v) \geq cover(T_v) + 1$*. We add the basic cover and leave a unit on the new leaf* $v$*. Or:*

2. **Leaves-Stems closed:** *In this case we use the budget inside* $T_v$ *to cover* $T_v$*.* No unit of credit is needed to be left on the new leaf $v$. *Recurse on* $T/T_v$ *(*$T$ *with* $T_v$ *contracted).*

*And we can tell between the two cases in linear time.*

The algorithm uses the lemma as follows:

Algorithm $Cover(T, E)$

1. $Q \leftarrow \emptyset$; /* This is the output solution */

2. $T' \leftarrow T$ ($T'$ is what remains)

3. Find a matching $\tilde{M}$ as described above

4. Add the twin links to derive Assumption 2.

5. **While** $T' \neq r$ **do:**

   (a) Find a semi-closed tree $T_v$

   (b) $Q \leftarrow Q \cup B(T_v)$ (The Basic cover is added to $Q$)

   (c) $T' \leftarrow T'/B(T_v)$ ($T_v$ is contracted)

   (d) If it is an extra-credit tree, find a new semi-closed tree $T_v$ /* No recursion here */

   (e) Else (leaves-stems closed tree $T_v$) $T_v$ recurse on $T/T_v$

6. Return $Q$

As long as it is an extra-credit tree $T_v$, we are able to add the basic cover and leave a unit of the lower bound on the new leaf $v$. This keeps all invariants.

A second possibility is that the tree $T_v$ is leaves-stems closed trees $T_v$. It is shown below that the links used to cover leaves and stems in $T_v$ and $T/T_v$ ($T$ with $T_v$ contracted) are disjoint.

**Theorem 30.** *TAP admits a* $4/3$ *approximation algorithm that runs in* $O(m \cdot \sqrt{n})$ *time*

*Proof.* If there is no recursion in the process, then we keep covering trees without overspending the budget, because all the tree are extra credit trees $\{T_v\}$. Namely, we add the basic cover and always are able to leave an *unspent* unit of the lower bound at the new leaf $v$. Note that this keeps all invariants. The new leaf is unmatched and own a unit of the lower bound. This covers $T$ with at most $4|F|/3$ links, because we do not overspend.

Else, we prove this by induction on the number of edges in the tree. Let $T_v$ be the first leaf-stem closed subtree. Let $F_v$ be the links used by $F$ to cover leaves and stems inside $T_v$. Since there is budget $4|M_v|/3 + |U_v| + gt(M_v)$ inside $T_v$ and we do not overspend, we have covered $T_v$ with at most $4 \cdot |F_v|/3$ links. Let $F'$ be the links used by $F$ to cover leaves and stems in $T/T_v$ ($T$ with $T_v$ contracted). Since the tree is leaf-stem closed, $F_v \cap F' = \emptyset$. Indeed, the lower bound is based *only on the links touching leaves and stems.*

By the induction hypothesis, we cover $T/T_v$ by $4|F'|3$ links. The total number of links is bounded by $4|F_v|/3 + 4|F'|/3 = 4|F|/3$. □

# 7 Proof of the main lemma

We prove the lemma by case analysis on the number of leaves and other properties.

## 7.1 Only matched pairs

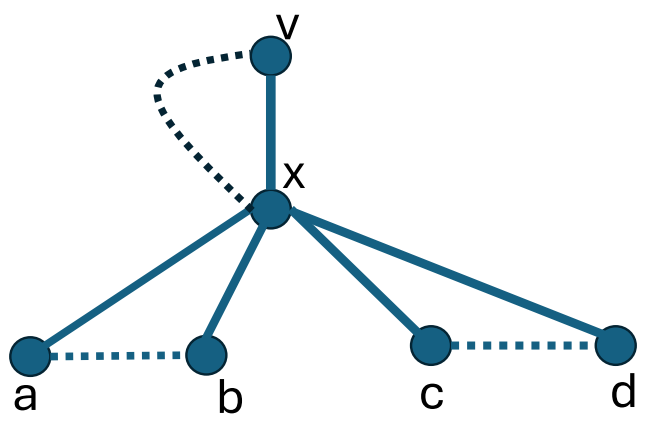

Figure 6: All leaves are matched

We first assume all leaves are matched. The assumption $|M_v| \leq 2$ implies $L_v \leq 4$. Since $L_v$ is even, either $|L_v| = 2$ or $|L_v| = 4$

**Claim 31.** $|L_v| = 4$

*Proof.* $|L_v| = 2$ implies the tree is a stem tree. This contradicts Claim. □

**Claim 32.** *The contraction of both links creates a unique leaf*

*Proof.* Otherwise, there is an additional leaf. It cannot be matched as this gives $|M_v| \geq 3$. It can also not be unmatched, as this contradicts our assumption. The claim follows. □

Thus, the tree is very specific if there are no unmatched leaves. It has the shape similar to the one of Figure 6.

**Claim 33.** *The tree is leaves-stems closed*

*Proof.* For the proof consult Figure 6. $x$ was a leaf in $T_v/M$ and $T_v/M$ is leaf-closed. □

From now on we assume there is at least one unmatched leaf.

## 7.2 Only unmatched leaves

**Claim 34.** *The tree is leaves-stems closed*

*Proof.* By Claim 28.1 the tree is stems=twins-closed. It is also closed for for the unmatched leaf. □

Note that leaf $c$ might be compound. And might contain many stems. Since the tree is closed for unmatched leaves, it is still stems-closed.

# 8 $|M_v| = 1$

## 8.1 $|L_v| = 3$

**Claim 35.**

1. *In case there are no stems the tree is an extra credit tree*

2. *In case there is an activated stem in $M$, the tree is a leaves-stems closed tree*

*Proof.* Consider Item 1. Let $ab$ be the link. This case meets the demands from Part B) left of Figure 4. Namely, the tree is $T_c$ closed and has no stems. By Claim 24, $w(ab) = 4/3 + 3/3 = 2$. We can use a unit to be left at $v$. Consider Item 2. Since the tree is closed for unmatched leaves, the claim follows from Claim 28.1. □

## 8.2 Four or more leaves

The following is proved like Claim 8.1, using Claim 25 instead of 24.

**Claim 36.** *If there is no activated stem the tree is an extra credit tree. If there is an activates stem, the tree is a leaves-stems closed tree.*

# 9 $|M_v| = 2$

First assume there is a stem that covers the root $v$. Namely, an $M_F$-activated stem. See figure 7.

**Claim 37.** *In this case the tree is an extra credit tree*

*Proof.* A simple calculation shows $|M_F| = 3$ cannot happen unless $M_F$ has two activated stems. If $|M_F(v)| = 3$, and $M_F(v)$ has only one twin link, the weight of $M_v$ can be decreased. If $|M_F| = 1$ or $|M_F(v)| = 3$ and $M_F(v)$ has two twin links, $w(M_F) > w(M)$. This gives an extra 1/3 in the lower bound. □

For the rest of the paper we assume there is no activated by $M_F$ stem and so, for every stem, $deg_F(s) = 0$.

We need the following definition:

**Definition 38.** *A tree* can be opened *if $M_v$ can be changed to another matching $M'$ of size* 2 *so that $W(M') = w(M)$ and $T_v$ is open for one of $a, b, c, d$.*

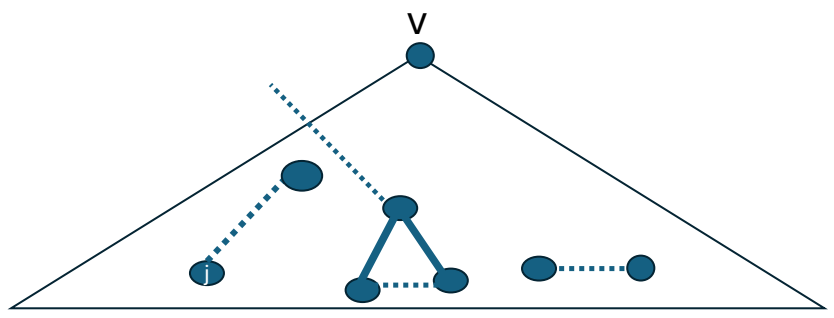

Figure 7: A stem covers $v$

Because the size of our matching is 2 we can always check if a tree can be opened. Since $a$ or $b$ or $c$ or $d$ is always an endpoint in any matching (the rest of the leaves share no links because of the independence invariant) searching for all possible matchings ca be done in linear time.

**Claim 39.** *We can reduce to the case $T_v$ cannot be opened*

*Proof.* A change in the matching so that, say, $a$ is $T_v$ open, implies that any strict subtree $T_u$ of $T_v$ cannot contain $a$. This implies that $|M_u| = 1$ and $T_u$ can be covered as seen above. Else, the tree strictly contains $T_v$. Since the tree $T$ is finite, we may assume that $T_v$ cannot be opened by opening a tree as many times as possible. □

**Claim 40.** *The tree is an extra credit tree*

*Proof.* Either, $w(M_F(v)) > w(M_v)$. In this case the claim is clear. Else, the unmatched leaves in $M_F$, namely $L_V \setminus V(M_F(v))$ cannot be $T_v$-open. Otherwise the tree can be opened. Thus, no leaf can cover the root. Also, $M_F$ has no activated stem. Thus, a non-leaf non-stem node covers the root. □

# 10 LP interpretation of parts of the algorithm: can the algorithm be used to prove integrality gap?

We give an LP interpretation to the independence invariant. After computing the matching, leaves are active sets (in the sense of [25]). We grow the dual variables. After they reach value 2 they meet. The dual can pay for the link (1/2 each) and the two active sets join into one (in the sense of [25]).

A similar interpretation can be given to covering a semi-closed tree. The dual variables of unmatched leaves grow to 1 and pay for the up edges. We take up edges since they are *maximal.* Unlike [25] there is no disjointness and a unit of the lower bound is placed at the new leaf.

Golden tickets can be described as *valid inequalities.* For example, we can set a variable for the links touching the stem and add an inequality saying that there sum is at least the value of the twin link. Similar valid inequalities can be added in case of other golden tickets. But, it is unclear if they make the integrality gap tighter.

## Acknowledgment

The author acknowledges the role Nutov has played in the motivation of the completion of the paper.

# A Reduction and run time

Since the largest weight is 5/3 by multiplying the weights by 3 the maximum weight becomes 5 and by duplicating vertices and links, the maximum weight becomes 1.

To reduce the problem of saturating leaves and stems to minimum cost perfect matching, we need to make sure the $v(i)$ nodes can be linked at weight 0. We link all $v(o)^s, v(1)^s$ of every stem . However, this might not be enough for a perfect matching to exist, since every stem whose twins are both matched outside can give a $v^s0)$ or $v^s(1)$ node that is unmatched. This quantity is not known. If the total quantity of the unmatched $v^s(i)$ is odd, we need an additional node $v^*$ joined to all $v^s(i)$ nodes with links, so that a perfect matching exists. Since this unknown quantity might be even, we need to run the algorithm twice, once with $v^*$ and one without, and take the solution with least weight.

**Identifying golden tickets:** This algorithm of [40] finds separating vertices. The trees of figure 4 has $v$ separating $T_c$ and $T_d$ its parent $p(v)$. It is also needed to check the tree is maximal. The recursion takes care of that. If a larger tree is bad, this tree is discarded. Separation of $T_c$ means that no node in $T_c$ reaches the parent $p(v)$ of $v$. Thus, if there is only $ab$ a twin link, $v$ has to separate all the other nodes, from its parent $p(v)$. The last case is that there are are two matched pairs. If there is also an $M$-activated twin link, then as discussed the tree is an extra credit tree. If no such link exists, then $T_v$ must separate all but 4 of its children and the other four have a perfect t matching.

All this can be done in time $O(m+n)$ [40].

**Sub-additivity:** Note that the algorithm of [34], [45] can be run $\Omega(n)$ times because of recursive calls for leaf-closed trees. However, the function $m \cdot \sqrt{n}$ is sub-additive. Namely, a function such that $f(a) + f(b) \leq f(a+b)$. See [33]. Hence, the worst case is when only a single matching is computed and the run time $O(m \cdot \sqrt{n})$ follows.

**Remark: The algorithm is fit for industrial use.** It might be lesser known in theory circles that [19] had been implemented by V. Kolmogorov's [29] and is used by Microsoft, IBM, Meta (Facebook), Intel, Google and various other companies. This means the algorithm for TAP *runs well in practice* since aside from finding a matching, we use the linear algorithm of [40].

# B Discussion

**Can the ratio be improved?** The case of 3 leaves is tight for the algorithm here. This *provably* implies that to improve the 4/3 for TAP new ideas are required. One idea is matching far away leaves first. If $T_c$ is large, this might fail.

The problem is already APX-hard if the links are between leaves [30]. The reduction is from the 3-dimensional Matching Problem (see [30].) The hardness given in [5] is explicit. 3-dimensional Matching cannot be approximated within better than $98/97 - \epsilon$. The reduction of [30] looses around a factor of 10. This makes LTL-TAP hard to approximate within around 901/900. The best known approximation algorithm for the 3 dimensional matching is $\frac{4}{3} + \varepsilon$ for any fixed $\varepsilon > 0$, [17]. The algorithm employs local search. This is similar to the *relative greedy technique*. See [43]. Both the ratio and technique have some similarities to [44]. The 4/3 might be a coincidence.